\documentclass[9pt,twocolumn,twoside]{osajnl}

\journal{ol} 

\setboolean{shortarticle}{true}

\title{On-chip Rotated Polarization Directional Coupler Fabricated by Femtosecond Laser Direct Writing}

\author[1,2]{Ci-Yu Wang}
\author[1,2]{Jun Gao}
\author[1,2,*]{Xian-Min Jin}

\affil[1]{State Key Laboratory of Advanced Optical Communication Systems and Networks, School of Physics and Astronomy, Shanghai Jiao Tong University, Shanghai 200240, China}
\affil[2]{Synergetic Innovation Center of Quantum Information and Quantum Physics, University of Science and Technology of China, Hefei, Anhui 230026, China}

\affil[*]{Corresponding author: xianmin.jin@sjtu.edu.cn}

\begin{abstract}
We present a rotated polarization directional coupler (RPDC) on a photonic chip. We demonstrate a double-track approach to modify the distribution of refractive index between adjacent tracks and form a single waveguide with arbitrary birefringent optical axis. We construct a RPDC with the two axis-rotated waveguides coupled in a strong regime. The obtained extinction ratios on average are about 16dB and 20dB for the corresponding orthogonal polarizations. We perform the reconstruction of Stokes vector to test the projection performance of our RPDC, and observe the average fidelities up to 98.1$\%$ and 96.0$\%$ for the perfectly initialized states in 0$^\circ$ and 45$^\circ$ RPDCs respectively.
\end{abstract}

\ociscodes{(130.0130) Integrated optics; (270.0270) Quantum optics; (270.5565) Quantum communications; (060.4510) Optical communications.}
\setboolean{displaycopyright}{true}

\begin{document}
\maketitle

Polarization manipulation of photons is a crucial approach in optical information processing, quantum communication and quantum computing. Especially in free space, there are many mature bulk optical devices for the preparation, control and projection of polarized states in very high precision. However, the scalability, robustness, cost and insertion loss have activated developing integrated polarization devices for large-scale applications \cite{Politi2008, Tanzilli2012, Brien2013, Metcalf2014, Sibson2017, Flamini2018}.

Femtosecond laser direct writing \cite{Davis1996, Minoshima2002, Kowalevicz2005, Eaton2005, Osellame2005, Eaton2008, Shane2011, osellame2012femtosecond, Arriola2013}, taking advantage of precise control of highly localized material modifications through nonlinear absorption processes, has been an emerging technique for controlling birefringence \cite{Kapron1972, Bricchi2004, Bhardwaj2004, Fernandes2012, Chen2014, McMillen2015} and constructing on-chip polarization devices. Moreover, its single-step, mask-free, phase-stable and cost-effective features have promised a powerful polarization-based technique for integrated photonics, such as polarization-dependent light attenuator \cite{Fangteng2013}, polarization-insensitive directional couplers \cite{Sansoni2012, Pitsios2017, Corrielli2018, Wang2018} (analogous to bulk optical beam splitters), polarization directional couplers \cite{Crespi2011, Fernandes20112, Dyakonov2017}, birefringent retarders, wave plates \cite{Fernandes20111, Heilmann2014, Corrielli2014} and integrated source \cite{Atzeni2018}. However, the orientation of these polarization devices is often fixed at horizontal or vertical with the standard configuration of femtosecond laser direct writing. 

\begin{figure}[hbt]
	\centering
	\fbox{\includegraphics[width=\linewidth]{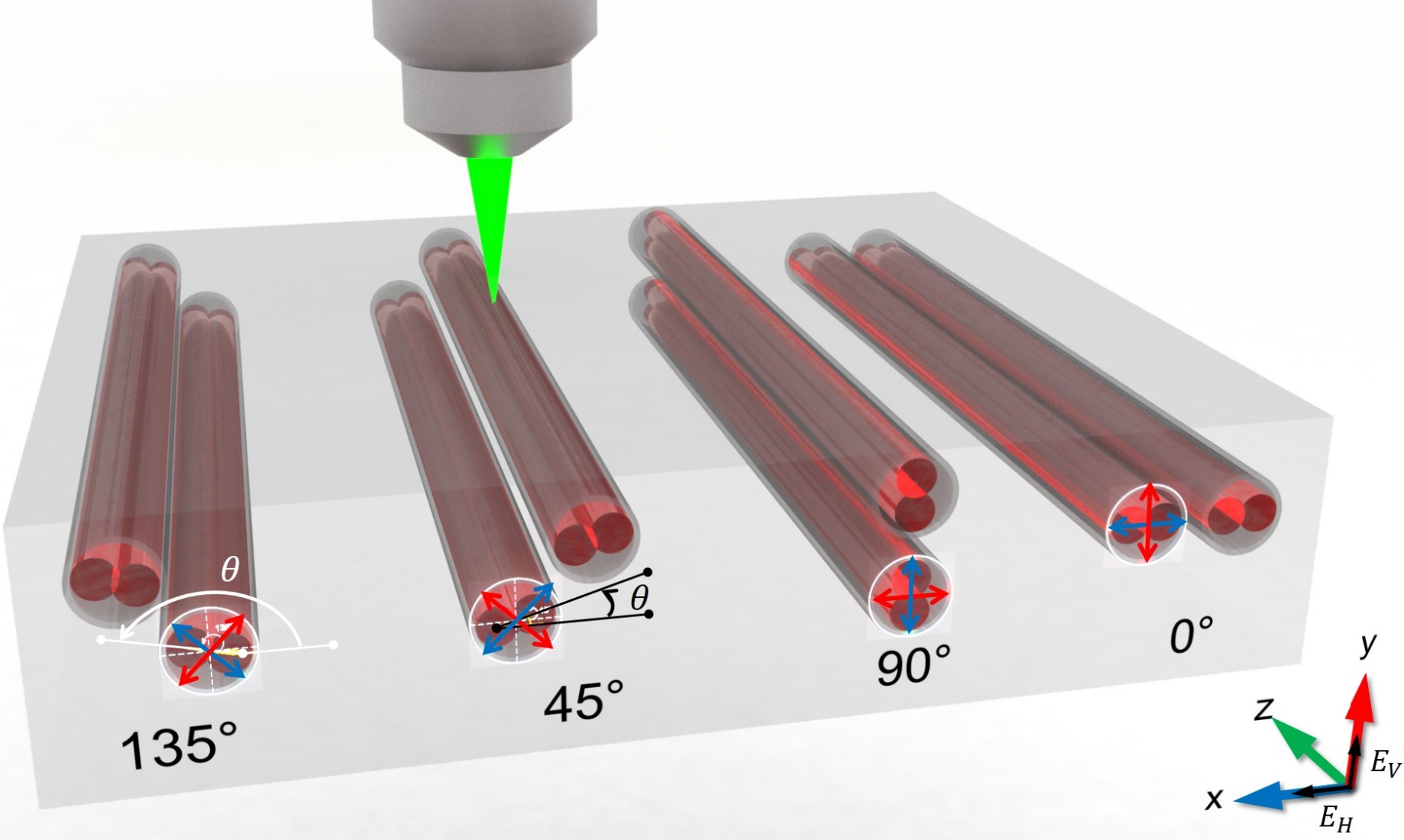}}
	\caption{Illustration of the fabricating process for double-track waveguides and sketch of the coupling cross-section of RPDCs. Each waveguide (grey) is composed of two adjacent tracks (dark red) with a geometrically radial and azimuthal ($\theta$) offset. The two waveguide arms are located parallel to the fast axis (blue lines with arrows at both ends) or orthogonally to the slow axis (red lines with arrows at both ends) to enable evanescent couplings.}
	\label{fig1}
\end{figure}

In this letter, we present an on-chip RPDC that can be constructed at arbitrary orientation by using a double-track approach. We adjust the relative radial and azimuthal position of the adjacent tracks to artificially set the birefringent optical axis of a single-mode waveguide with a high transmittance. With the two axis-rotated waveguides, we can construct a RPDC along the same rotated axis, with which we are able to make the projection measurement on any pair of orthogonal basis rather than either horizontal or vertical. 

As is shown in Fig. \ref{fig1}, the waveguide (grey) is composed of two adjacent parallel tracks (dark red) with a specific offset (radial $\&$ azimuthal). Different from the method of defect tracks \cite{Heilmann2014} written nearby, our track pairs are laid out quite close with a bit overlapped. Therefore, the two tracks here are able to transmit light simultaneously without energy dissipation into the defect track. An optimal radial separation can be found where the track pairs are able to guide light behaving as a single-mode waveguide \cite{Jovanovic2010}. Moreover, a precise change of relative position in the track pairs may induce the anisotropic and inhomogeneous distribution of the refractive index, generating an artificial rotation in birefringent optical axis.

We employ a regenerative amplifier based on Yb:KGW lasing medium, with a pulse duration of 290 fs, a repetition rate of 1 MHz and a central wavelength of 513 nm, to conduct the fabrication. A 100$\times$ (0.70NA) microscope objective is adopted to tightly focus the subsequent linearly polarized (along the writing direction) laser pulses into a depth of 170 $\mu$m underneath a commercially available fused silica substrate surface. We move the substrate and laser focal spot with three air-bearing translation stages, and the formed trajectory of the focal spot is the track. The writing speed in our experiment is about 1.268 mm/s with a resolution of 0.1 $\mu$m. We fix the writing laser at an optimized pulse energy of 180nJ and shape the beam by using a cylindrical lens \cite{Osellame2003} with a focal length of 70 cm.

The single-track is able to guide a single mode at 780 nm as well as the double-track waveguide. However, its birefringence is around the order of $10^{-5} \sim 10^{-6}$, leaving the adjacent track to tenfold stress the birefringence \cite{LAFernandes2013}. Also, the single-track waveguide modes are generally larger compared to double-track waveguide due to decreased index change at lower fluence \cite{Shah2005}.

\begin{figure}[h]
	\centering
	\fbox{\includegraphics[width=\linewidth]{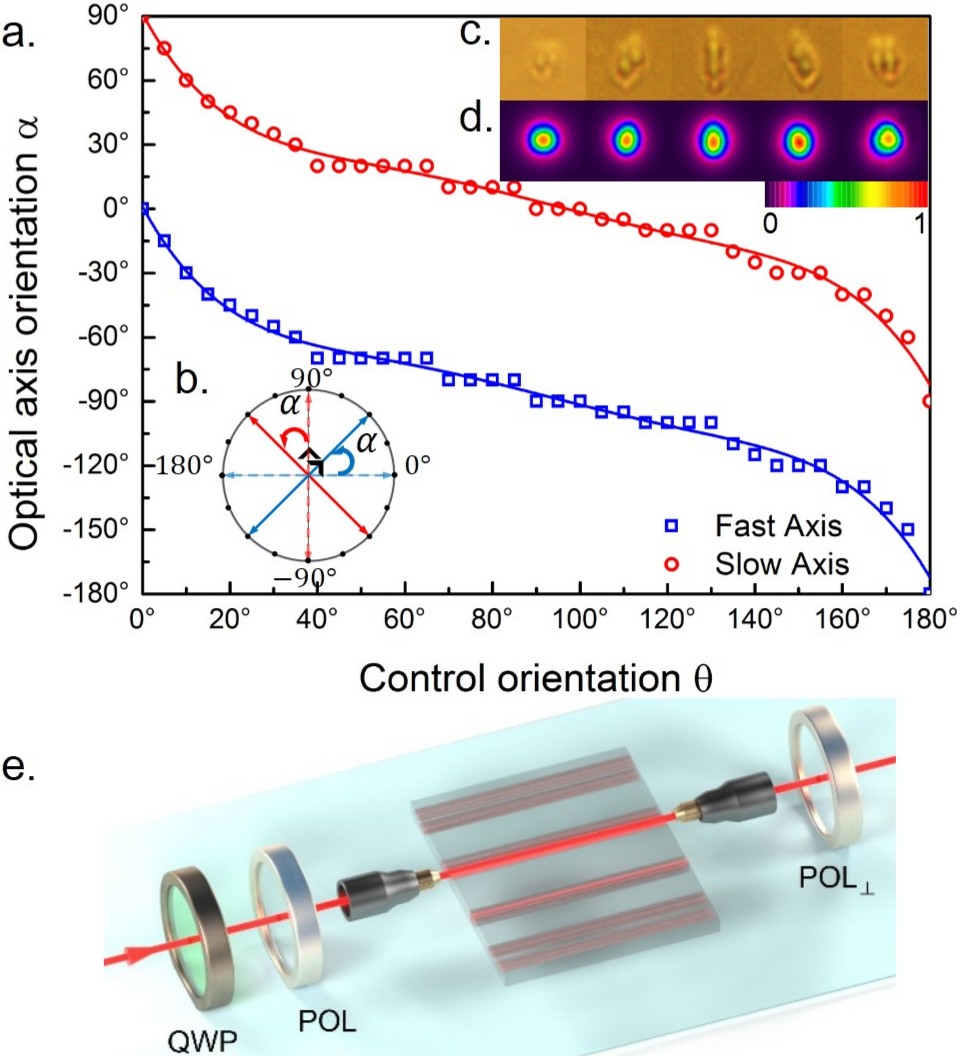}}
	\caption{a. Experimental results and best fit model of the artificial birefringence optical axis reorientation $\alpha$ as a function of the azimuthal offset control orientation $\theta$. b. Schematic of rotation plane of the two rotated orthogonal optical axis. c. Morphology of the waveguide cross-sections presented by microscope end-view of track pairs fabricated with 100 $\times$-objective, 2 $\mu$m radial offset in fused silica substrate. d. Corresponding near-field output modes of waveguides with different azimuthal offsets. e. Setup for birefringent optical axis confirmation. POL: polarizer, QWP: quarter wave plate.}
	\label{fig2}
\end{figure}

Fig. \ref{fig2}a shows the induced rotation of the birefringent optical axis ($\alpha$) dependent on the azimuthal orientation angles ($\theta$) of the track pairs from 0$^\circ$ to 180$^\circ$ per 5$^\circ$ at a fixed optimized radial separation of 2 $\mu$m. We deduce the orientation of birefringent optical axis of the waveguide by placing two crossed polarizers (Pol and Pol$_{\perp}$) before and after the chip (Fig. \ref{fig2}e) \cite{Yang2004,Corrielli2014,Heilmann2014}. The minimum transmission reveals the orientation of the birefringent optical axis, and the maximum is always 45$^\circ$ away from the minimum. Although not linearly proportional, the fast and slow axis are confirmed as orthogonal and tunable in the plane marked in Fig. \ref{fig2}b by controlling the orientation angles ($\theta$), behaving as a stable and controllable transmission of $\alpha$-polarized or the corresponding orthogonal polarized light through the chip. The end-view microscope images (correspond to an average area of 5.6 $\mu$m $\times$5.8 $\mu$m) of waveguides cross-section composed of the track pairs can be found in Fig. \ref{fig2}c, for the real pictures of changing applied writing parameters of control orientation ($\theta$) and radial separation. Moreover, we find that the near-field profiles of the guided modes (Fig. \ref{fig2}d) (correspond to an average area of 6.8 $\mu$m $\times$6.9 $\mu$m width) at 780nm wavelength preserve well with a comparably small influence induced by differential $\theta$. We also observe a high average total transmission rate up to 60 $\%$ through 19.64mm waveguide propagation with a 60$\times$ objective in free space, and over 50$\%$ transmission coupled into single-mode fiber.

\begin{figure}[h]
	\centering
	\fbox{\includegraphics[width=\linewidth]{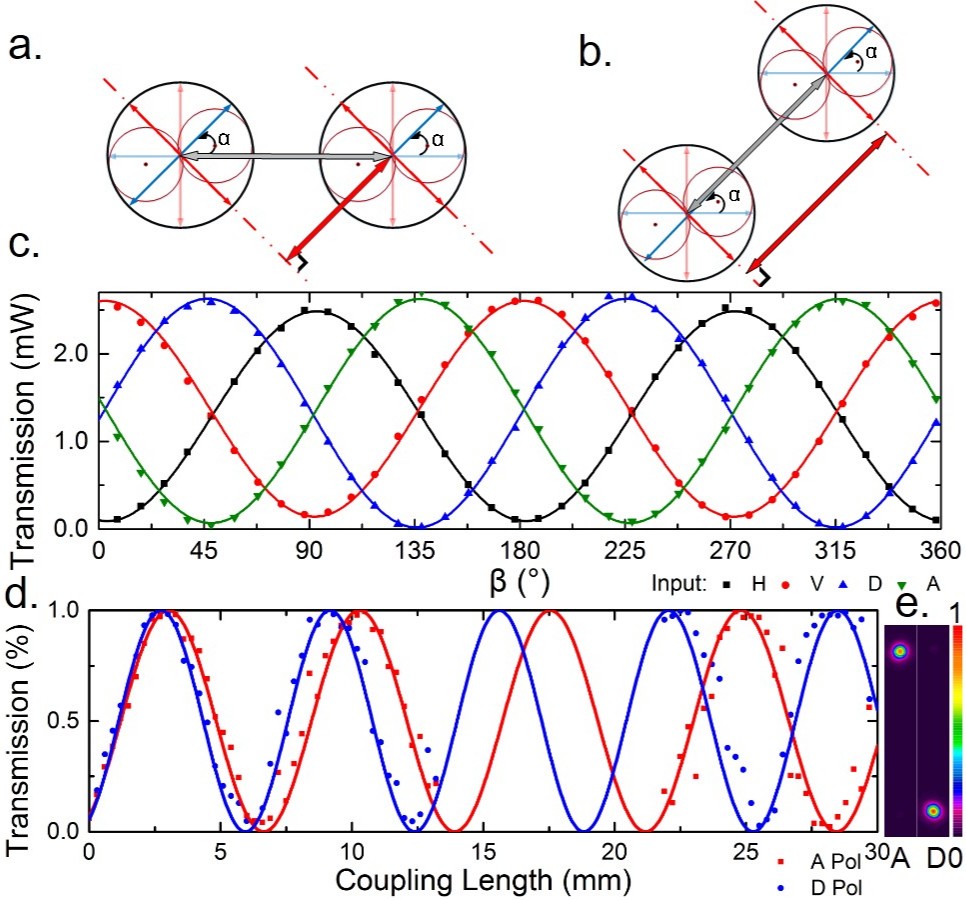}}
	\caption{a. The configuration of two waveguides located in a plane of same depth. b. The configuration of two waveguides located with a rotation of $\alpha$ from the plane. c. Polarization analysis of the 45$^\circ$ rotated parallel coupling region with different linearly input state (H, V, D, A Pol.). d. Normalized transmission power of anti-diagonal ($135^{\circ}$, A Pol.) and diagonal ($45^{\circ}$, D Pol.) polarized light through a $45^{\circ}$ RPDC. e. Extinction performance of the 45$^\circ$ RPDC.}
	\label{fig3}
\end{figure}

In order to achieve a polarization directional coupler, one need to utilize the difference of coupling coefficients between the input light that polarized parallel and perpendicular to the birefringent optical axis. It has been proved that the coupling constants for H and V polarized light vary from each other, not only caused by projection magnitude, but also due to anisotropy under different spatial dispositions of waveguide arms with fixed horizontal and vertical optical axis \cite{Szameit2007}. Also, a link between the geometry of the waveguides, the polarization of light and the coupling constants has been revealed \cite{Rojas2014}. The wanted difference of evanescent coupling may become weak and complex if the two waveguides are kept in the same plane while their birefringent optical axis are rotated (Fig. \ref{fig3}a). This enlightens us to form a 3D tilted coupler with feasible experimental modified configuration to produce a compact and balanced RPDC with rotated optical axis ( Fig. \ref{fig3}b).  Once the birefringent optical axis is confirmed, the evanescent light coupling of the two waveguides should still subject to the coupling mode theory \cite{Snyder1986}, in which the field amplitude evolution $a_1(z)$ and $a_1(z)$ in two waveguide arms follows the equation:
\begin{equation}
\left\{
	\begin{array}{lr}
	\frac{\mathsf{d}a_1}{\mathsf{d}z} = -i\beta_1(z)a_1 + K_{12} a_2 , & \\
	\frac{\mathsf{d}a_2}{\mathsf{d}z} = K_{21}a_1 - i\beta_2(z)a_2 . &
	\end{array}
\right.
\end{equation}

In the ideal case, we take the wavenumber dependent on the propagation coordinate $\beta_1 = \beta_2$, and the coupling coefficient $K_{12} = K_{21}$ considering the modification of the mode's effective refractive indices are identical. It is not difficult to retrieve the shortest coupling length reaching a polarization directional coupler remains $\pi/2(K_S-K_F)$, where $K_S$ and $K_F$ are the two coupling coefficients for the two orthogonal polarization states, supposing $K_S>K_F$ . 

The coupling ratio (field amplitudes) with the configuration shown in Fig. \ref{fig3}b can be calculated as:
\begin{equation}
\left\{
	\begin{array}{lr}
	T = sin(KZ), & \\
	R = \sqrt{1 - T^2}. &
	\end{array}
\right.
\end{equation}

\begin{figure}[t]
	\centering
	\fbox{\includegraphics[width=\linewidth]{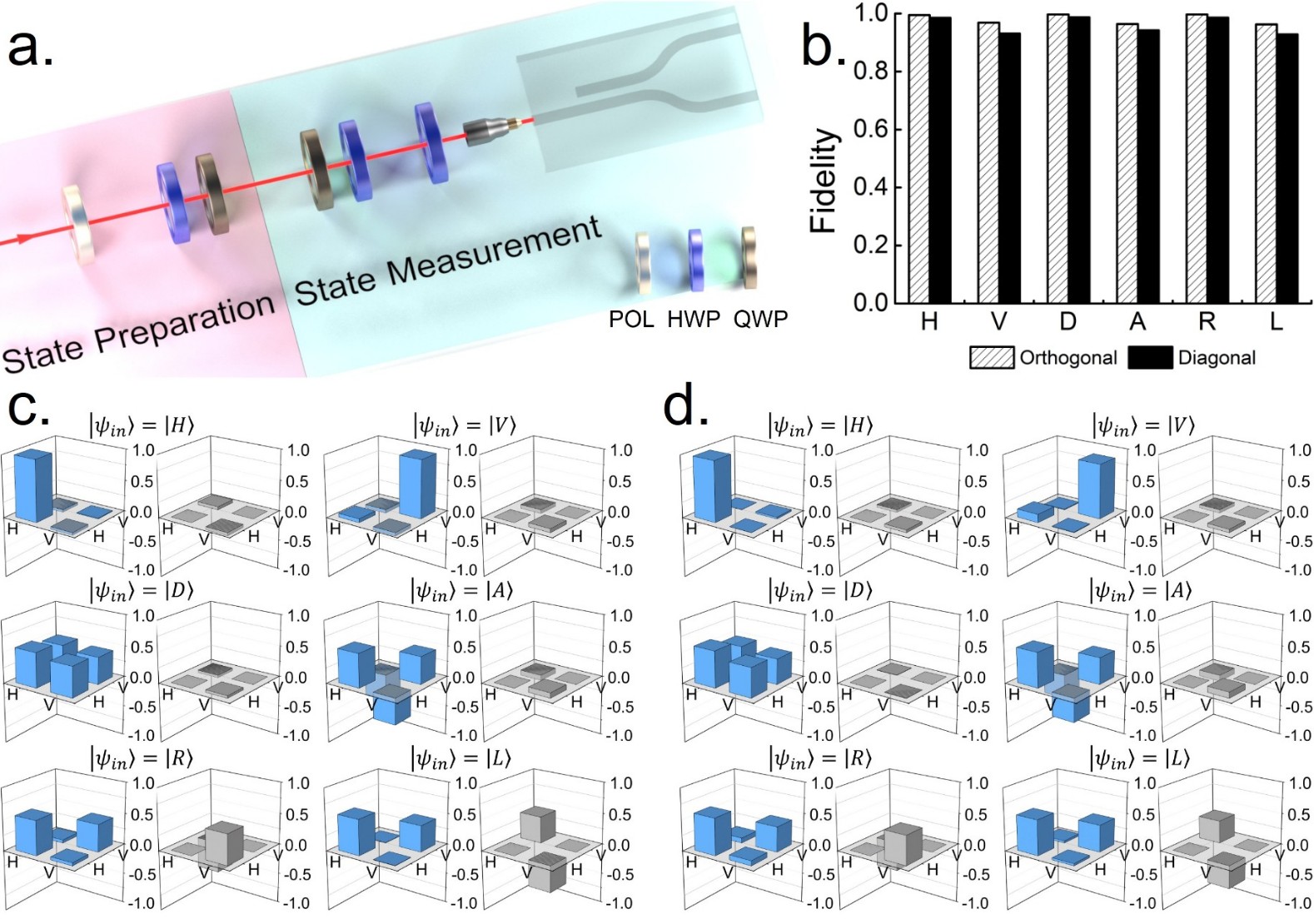}}
	\caption{a. Schematic of experimental setup for reconstruction of Stokes vector. POL: Polarizer HWP: half wave plate QWP: quarter wave plate b. The measured fidelities of different initialized states with $0^{\circ}$ and $45^{\circ}$ RPDCs. c. Reconstructed density matrix obtained with $0^{\circ}$ RPDC. d. Reconstructed density matrix obtained with $45^{\circ}$ RPDC. The blue parts represents the real components and grey parts for imaginary components.}
	\label{fig4}
\end{figure}

To validate this theory, we fabricate the second double-tracks waveguide arm 7 $\mu$m beside (parallel to the fast axis) to enable the evanescent light coupling shown in Fig. \ref{fig3}b. In order to exclude potential additional birefringent stress that the second waveguide arm may cause to rotate optical axis, we made polarization analysis of the 45° rotated birefringent optical axis of the parallel straight coupling region by two crossed polarizers as is mentioned above. As is shown in \ref{fig3}c, the linearly input H polarized light is measured as V polarized output light, etc. The input D/A polarized light travel along the optical axis remains visibility as high as $98\% \sim 99\%$. Transmission power of the 45$^\circ$ RPDC is characterized by different coupling lengths as is shown in Fig. \ref{fig3}d. We calculate the ratio with beam diameter definition of full-width, half max. From our experiment, the coupling lengths reaching the 0$^\circ$ and 45$^\circ$ RPDCs are equal, which well verifies the theory of our RPDC. The two waveguide arms are split adiabatically through a "S curve" as is shown in Fig. \ref{fig4}a. It should be noticed that evanescent coupling may also occur during the bending region and a phase term should be added in the argument of the sine function. Therefore, the parallel coupling length reaching a RPDC therefore could be shortened from 28.5mm to 23mm. Experimentally, we observe average extinction ratios up to 16dB and 20dB (Fig. \ref{fig3}e) for the diagonal (D) and anti-diagonal (A) polarizations, respectively. Here we define extinction ratio as:
\begin{equation}
\left\{
\begin{array}{lr}
ER_T = |10\log\frac{T_F}{T_S}| \quad (dB) , & \\
ER_R = |10\log\frac{R_F}{R_S}| \quad (dB) . &
\end{array}
\right.
\end{equation}
where $p_{i=1,2}$ is the output power of different waveguide direction coupler output arms respectively.

To further test the quality of the fabricated RPDC, we perform the reconstruction of Stokes vector by using simple linear tomography \cite{Stokes1851,James2001} technique shown in Fig. \ref{fig4}a. In our experiment, we initialize high-purity polarization states with high-extinction polarizer and wave plates at the state preparation part. In the state measurement part, we set the QWP and HWP to map the states to horizontal/vertical basis. Unlike the standard bulk optical implementation, the projection measurements are carried out by a RPDC rather than a polarization beam splitter. The additional HWP rotate the frame to our RPDC. In this way, we follow the standard procedure to perform state tomography while our RPDC acts as a polarization beam splitter to make projection measurements. The density matrix for the polarization degrees of light can be related to the Stokes parameters by the formula:
\begin{equation}
	\hat{\rho} = \frac{1}{2}\sum\limits_{i=0}^3\frac{S_i}{S_0}\hat{\sigma_i}
\end{equation}
where $\hat{\sigma_{i=0}}$ is the identity operator and $\hat{\sigma}_{i=1,2,3}$ are the Pauli operators. 
\begin{equation}
\left\{
\begin{array}{lr}
S_0 = P_{\left| 0 \right\rangle} + P_{\left| 1 \right\rangle}, & \\
S_1 = P_{\frac{1}{\sqrt{2}}(\left| 0 \right\rangle + \left| 1 \right\rangle)} - P_{\frac{1}{\sqrt{2}}(\left| 0 \right\rangle - \left| 1 \right\rangle)}, & \\
S_2 = P_{\frac{1}{\sqrt{2}}(\left| 0 \right\rangle + i\left| 1 \right\rangle)} - P_{\frac{1}{\sqrt{2}}(\left| 0 \right\rangle - i\left| 1 \right\rangle)}, & \\
S_3 = P_{\left| 0 \right\rangle} - P_{\left| 1 \right\rangle}.
\end{array}
\right.
\end{equation} 
where $P_{\left| \phi \right\rangle}$ represents the probability to measure the state $\phi$.

Ideally, the fidelities measured with a perfect RPDC should be 100 $\%$. Here, we observe the average fidelities up to 98.1$\%$ and 96.0$\%$ in 0$^\circ$ and 45$^\circ$ RPDCs respectively, where we rigorously perform a maximum likelihood estimation to keep density matrix physical. The detailed reconstructed density matrix are shown in Fig. \ref{fig4}c and \ref{fig4}d with error bars omitted as they are too small to be visible. The projection performance of RPDCs with no much difference identifies the ability to make general polarization projections on a photonic chip.

In conclusion, a new double-track approach is presented here to realize waveguide with rotated birefringent optical axis and to construct an on-chip RPDC with the two axis-rotated waveguides coupled in a strong regime. We test the quality of the on-chip RPDC by performing the reconstruction of Stokes vector with perfectly initialized polarization states. The features of straightforward fabrication, straightforward polarization projection and high transmittance make this on-chip RPDC a good candidate for large-scale polarization-based optical communication and quantum information processing.

\section*{Acknowledgments}
The authors thank Roberto Osellame and Jian-Wei Pan for helpful discussions. This work was supported by National Key R\&D Program of China (2017YFA0303700); National Natural Science Foundation of China (NSFC) (61734005, 11761141014, 11690033); Science and Technology Commission of Shanghai Municipality (STCSM) (15QA1402200, 16JC1400405, 17JC1400403); Shanghai Municipal Education Commission (SMEC)(16SG09, 2017-01-07-00-02-E00049); X.-M.J. acknowledges support from the National Young 1000 Talents Plan..

\bibliography{RPDC}

\bibliographyfullrefs{RPDC}

\ifthenelse{\equal{\journalref}{aop}}{%
\section*{Author Biographies}
\begingroup
\setlength\intextsep{0pt}
\begin{minipage}[t][6.3cm][t]{1.0\textwidth} 
  \begin{wrapfigure}{L}{0.25\textwidth}
    \includegraphics[width=0.25\textwidth]{john_smith.eps}
  \end{wrapfigure}
  \noindent
  {\bfseries John Smith} received his BSc (Mathematics) in 2000 from The University of Maryland. His research interests include lasers and optics.
\end{minipage}
\begin{minipage}{1.0\textwidth}
  \begin{wrapfigure}{L}{0.25\textwidth}
    \includegraphics[width=0.25\textwidth]{alice_smith.eps}
  \end{wrapfigure}
  \noindent
  {\bfseries Alice Smith} also received her BSc (Mathematics) in 2000 from The University of Maryland. Her research interests also include lasers and optics.
\end{minipage}
\endgroup
}{}

\end{document}